\documentclass[12pt]{article}
\usepackage{epsfig}

\newcommand{\mste}{m_{\tilde{t}_1}}
\newcommand{\mstz}{m_{\tilde{t}_2}}

\newcommand{\msb}{m_{\tilde{b}}}

\newcommand{\At}{A_t}
\newcommand{\Ab}{A_b}

\newcommand{\msbar}{$\overline{\rm{MS}}$}

\def\order#1{${\cal O}(#1)$}
\newcommand{\cL}{{\cal L}}

\newcommand{\cp}{{\cal CP}}

\newcommand{\onel}{one-loop}

\newcommand{\fh}{{\em FeynHiggs}}

\newcommand{\MW}{M_W}
\newcommand{\MZ}{M_Z}
\newcommand{\MA}{M_A}

\newcommand{\Mh}{M_h}
\newcommand{\MH}{M_H}

\newcommand{\mt}{m_{t}}

\newcommand{\mgl}{m_{\tilde{g}}}

\newcommand{\Stop}{\tilde{t}}

\newcommand{\Stope}{\tilde{t}_1}
\newcommand{\Stopz}{\tilde{t}_2}

\newcommand{\tst}{\theta_{\tilde{t}}}

\newcommand{\tsf}{\theta\kern-.20em_{\tilde{f}}}
\newcommand{\tsfp}{\theta\kern-.20em_{\tilde{f}\prime}}
\newcommand{\tsq}{\theta\kern-.15em_{\tilde{q}}}
\newcommand{\sw}{s_W}
\newcommand{\cw}{c_W}
\newcommand{\sweff}{\sin^2\theta_{\mathrm{eff}}}
\newcommand{\cweff}{\cos^2\theta_{\mathrm{eff}}}

\newcommand{\sintt}{\sin\tst}

\newcommand{\VL}{\left( \begin{array}{c}}
\newcommand{\VR}{\end{array} \right)}
\newcommand{\ML}{\left( \begin{array}{cc}}
\newcommand{\MLd}{\left( \begin{array}{ccc}}
\newcommand{\MLv}{\left( \begin{array}{cccc}}
\newcommand{\MR}{\end{array} \right)}
\newcommand{\dd}{\partial}

\newcommand{\tb}{\tan \beta}

\newcommand{\gev}{\,\, \mathrm{GeV}}
\newcommand{\mev}{\,\, \mathrm{MeV}}

\newcommand{\BC}{\begin{center}}
\newcommand{\EC}{\end{center}}
\newcommand{\BE}{\begin{equation}}
\newcommand{\EE}{\end{equation}}
\newcommand{\BEA}{\begin{eqnarray}}
\newcommand{\BEAnn}{\begin{eqnarray*}}
\newcommand{\EEA}{\end{eqnarray}}
\newcommand{\EEAnn}{\end{eqnarray*}}

\newcommand{\id}{{\rm 1\kern-.12em
\rule{0.3pt}{1.5ex}\raisebox{0.0ex}{\rule{0.1em}{0.3pt}}}}
\newcommand{\lsim}
{\;\raisebox{-.3em}{$\stackrel{\displaystyle <}{\sim}$}\;}
\newcommand{\gsim}
{\;\raisebox{-.3em}{$\stackrel{\displaystyle >}{\sim}$}\;}

\newcommand{\gf}{G_F}

\def\al{\alpha}

\def\als{\alpha_s}

\def\de{\delta}

\def\si{\sigma}

\def\Ga{\Gamma}

\def\De{\Delta}

\newcommand{\SLASH}[2]{\makebox[#2ex][l]{$#1$}/}
\newcommand{\Dslash}{\SLASH{D}{.5}\,}
\newcommand{\dslash}{\SLASH{\dd}{.15}}

\newcommand{\epem}{$e^+e^-$}

\newcommand{\etal}{{\em et al.}}

\newcommand{\as}{\alpha_s}

\newcommand\pubnumber{UPR--924--T\\ BNL--HET--01/3}
\newcommand\pubdate{\today}
\newcommand\hepnumber{hep-ph/0102083}

\def\csumb{$^1$Department of Physics and Astronomy, University of Pennsylvania, \\
Philadelphia PA 19104--6396, USA

\vspace*{0.4cm}

$^2$BNL, Physics Department, Upton NY 11973, USA

\vspace*{0.4cm}
}

\def\Title#1{\begin{center} {\Large\bf #1 } \end{center}}
\def\Author#1{\begin{center}{ \sc #1} \end{center}}
\def\Address#1{\begin{center}{ \it #1} \end{center}}

\newcommand\pubblock{\rightline{\begin{tabular}{l} \pubnumber\\
         \pubdate\\ \hepnumber \end{tabular}}}
\newenvironment{Abstract}{\begin{quotation}  }{\end{quotation}}
\newenvironment{Presented}{\begin{quotation} \begin{center} 
             Presented by {\sc Jens Erler} at the\end{center}
      \begin{center}\begin{large}}{\end{large}\end{center} \end{quotation}}
\def\Acknowledgments{\bigskip  \bigskip \begin{center}
          \large\bf Acknowledgments\end{center}}

\makeatletter
\def\section{\@startsection{section}{0}{\z@}{5.5ex plus .5ex minus
 1.5ex}{2.3ex plus .2ex}{\large\bf}}
\def\subsection{\@startsection{subsection}{1}{\z@}{3.5ex plus .5ex minus
 1.5ex}{1.3ex plus .2ex}{\normalsize\bf}}
\def\subsubsection{\@startsection{subsubsection}{2}{\z@}{-3.5ex plus
-1ex minus  -.2ex}{2.3ex plus .2ex}{\normalsize\sl}}

\renewcommand{\@makecaption}[2]{%
   \vskip 10pt
   \setbox\@tempboxa\hbox{\small #1: #2}
   \ifdim \wd\@tempboxa >\hsize     
       \small #1: #2\par          
     \else                        
       \hbox to\hsize{\hfil\box\@tempboxa\hfil}
   \fi}

 \def\citenum#1{{\def\@cite##1##2{##1}\cite{#1}}}
 
\newcount\@tempcntc
\def\@citex[#1]#2{\if@filesw\immediate\write\@auxout{\string\citation{#2}}\fi
  \@tempcnta\z@\@tempcntb\m@ne\def\@citea{}\@cite{\@for\@citeb:=#2\do
    {\@ifundefined
       {b@\@citeb}{\@citeo\@tempcntb\m@ne\@citea\def\@citea{,}{\bf ?}\@warning
       {Citation `\@citeb' on page \thepage \space undefined}}%
    {\setbox\z@\hbox{\global\@tempcntc0\csname b@\@citeb\endcsname\relax}%
     \ifnum\@tempcntc=\z@ \@citeo\@tempcntb\m@ne
       \@citea\def\@citea{,}\hbox{\csname b@\@citeb\endcsname}%
     \else
      \advance\@tempcntb\@ne
      \ifnum\@tempcntb=\@tempcntc
      \else\advance\@tempcntb\m@ne\@citeo
      \@tempcnta\@tempcntc\@tempcntb\@tempcntc\fi\fi}}\@citeo}{#1}}
\def\@citeo{\ifnum\@tempcnta>\@tempcntb\else\@citea\def\@citea{,}%
  \ifnum\@tempcnta=\@tempcntb\the\@tempcnta\else
  {\advance\@tempcnta\@ne\ifnum\@tempcnta=\@tempcntb \else\def\@citea{--}\fi
    \advance\@tempcnta\m@ne\the\@tempcnta\@citea\the\@tempcntb}\fi\fi}
\makeatother

\def\beq{\begin{equation}}
\def\eeq#1{\label{#1}\end{equation}}
\def\eeqn{\end{equation}}

\newenvironment{Eqnarray}%
   {\arraycolsep 0.14em\begin{eqnarray}}{\end{eqnarray}}
\def\beqa{\begin{Eqnarray}}
\def\eeqa#1{\label{#1}\end{Eqnarray}}
\def\eeqan{\end{Eqnarray}}

\let\bar=\overbar

\def\etal{{\it et al.}}

\def\Dslash{\not{\hbox{\kern-4pt $D$}}}
\def\dslash{\not{\hbox{\kern-2pt $\del$}}}

\def\mt{m_t}

\def\msb{{\bar{\ssstyle M \kern -1pt S}}}

\def\lsim{\mathrel{\raise.3ex\hbox{$<$\kern-.75em\lower1ex\hbox{$\sim$}}}}
\def\gsim{\mathrel{\raise.3ex\hbox{$>$\kern-.75em\lower1ex\hbox{$\sim$}}}}

\begin{document}
\begin{titlepage}
\pubblock

\vfill
\def\thefootnote{\fnsymbol{footnote}}
\Title{GigaZ: High Precision Tests\\[.5em] of the SM and the MSSM}
\vfill
\Author{J.~Erler$^{1}$ and
S.~Heinemeyer$^{2}$}
\Address{\csumb}
\vfill
\begin{Abstract}
The high-energy \epem collider TESLA can be operated in the GigaZ mode on
the $Z$~resonance, producing \order{10^9} $Z$~bosons per year.
This will allow the measurement of the effective
electroweak mixing angle to an accuracy of 
$\de\sweff \approx \pm 1 \times 10^{-5}$. 
Similarly the $W$~boson mass is expected to be measurable with an error of 
$\de\MW \approx \pm 6$~MeV near the $W^+W^-$ threshold.  We discuss
the impact of these observables on the accuracy with which the Higgs
boson mass can be determined from  
loop corrections within the Standard Model. 
We also study indirect constraints on new mass scales
within the Minimal Supersymmetric Standard Model.
\end{Abstract}
\vfill
\begin{Presented}
5th International Symposium on Radiative Corrections \\ 
(RADCOR--2000) \\[4pt]
Carmel CA, USA, 11--15 September, 2000
\end{Presented}
\vfill
\end{titlepage}
\def\thefootnote{\arabic{footnote}}
\setcounter{footnote}{0}
%


\section{Introduction}

The high-energy \epem\ linear collider TESLA is being designed to operate on 
top of the $Z$~boson resonance by adding a bypass to the main beam line. 
Given the high luminosity, $\cL = 7 \times 10^{33} \rm{cm}^{-2} \rm{s}^{-1}$, 
and the cross section, $\si_Z \approx 30 \rm{~nb}$, about $2 \times 10^9~Z$
events can be generated in an operational year of $10^7 \rm{s}$. We will 
therefore refer to this option as the GigaZ mode of the machine. Moreover, by 
increasing the collider energy to the $W$-pair threshold, about $10^6$ $W$
bosons can be generated at the optimal energy point for measuring the $W$ boson
mass, $\MW$, near threshold and about $3 \times 10^6$ $W$ bosons at the energy 
of maximal cross section. The large increase in the number of $Z$~events by 
two orders of magnitude as compared to LEP1 and the increasing precision in 
the measurements of $W$~boson properties, open new opportunities for high 
precision physics in the electroweak sector~\cite{gigaz}.

By adopting the Blondel scheme~\cite{blondel}, the left-right asymmetry,
$A_{LR} \equiv 2 (1 - 4 \sweff)/(1 + (1 - 4 \sweff)^2)$,
can be measured with very high precision, 
$\de A_{LR} \approx \pm 10^{-4}$~\cite{moenig}, when both, electrons and 
positrons, are polarized longitudinally. From $A_{LR}$ the mixing angle 
in the effective leptonic vector coupling of the on-shell $Z$~boson,
$\sweff$, can be determined to an accuracy~\cite{moenig}, 
\BE
\de\sweff \approx \pm\,1 \times 10^{-5},
\EE
while the $W$~boson mass is expected to be measurable
within~\cite{wwthreshold} 
\BE
\label{delmw}
\de\MW \approx \pm\,6 \mev.
\EE

Besides the improvements in $\sweff$ and $\MW$, GigaZ has the potential 
to determine the total $Z$~width within $\de\Gamma_Z = \pm 1$~MeV; the ratio 
of hadronic to leptonic partial $Z$~widths with a relative uncertainty of 
$\de R_l/R_l = \pm 0.05 \%$; the ratio of the $b\bar{b}$ to the hadronic 
partial widths with a precision of $\de R_b = \pm 1.4 \times 10^{-4}$; and 
to improve the $b$ quark asymmetry parameter $A_b$ to a precision of
$\pm 1 \times 10^{-3}$~\cite{moenig}. These additional measurements offer 
complementary information on the Higgs boson mass, $\MH$, but also on 
the strong coupling constant, $\als$, which enters the radiative corrections 
in many places. This is desirable in its own right, and in the present context 
it is important to control $\as$ effects from higher order loop contributions 
to avoid confusion with Higgs effects. Indirectly, a well known $\as$ would 
also help to control $\mt$ effects, since $\mt$ from a threshold scan at 
a linear collider will be strongly correlated with $\as$. We find that via 
a precise measurement of $R_l$, GigaZ would provide a clean determination of 
$\as$ with small error, 
\BE
\de\as \approx \pm\, 0.001 , 
\label{delalphas}
\EE
and consequently a smaller uncertainty in $\de\mt$ compared to a linear 
collider, given identical threshold data ($5 \times 10\; {\rm fb}^{-1}$).
The anticipated precisions for the most relevant electroweak observables 
at the Tevatron (Run IIA and IIB), the LHC, a future linear collider, LC,
and GigaZ are summarized in Table \ref{tab:precallcoll}. 

\def\tableline{\noalign{
\hrule height.7pt depth0pt\vskip3pt}}
\begin{table}[t!]
\caption{\label{tab:precallcoll}
Expected precision at various colliders for $\sweff$, $\MW$, $\mt$ and 
the (lightest) Higgs boson mass, $\MH$, at the reference value $\MH = 110$~GeV.
Run IIA refers to 2~fb$^{-1}$ integrated luminosity per experiment collected 
at the Tevatron with the Main Injector, while Run IIB assumes the accumulation 
of 30~fb$^{-1}$ by each experiment. LC corresponds to a linear
collider without the GigaZ  
mode. (The entry in parentheses assumes a fixed target polarized M\o ller 
scattering experiment using the $e^-$ beam.) 
Concerning the present value of $\MW$ some improvement can be expected
from the final analysis of the LEP~2 data.
$\de\mt$ from the Tevatron and 
the LHC is the error for the top pole mass, while at the top threshold in 
\epem\ collisions the \msbar\ top-quark mass can be determined. The smaller
value of $\de\mt$ at GigaZ compared to the LC is due to the prospective
reduced uncertainty in $\alpha_s$, which affects the relation between
the mass parameter directly extracted at the top threshold and the \msbar\
top-quark mass.}
\begin{center}
\setlength{\tabcolsep}{9pt}
\renewcommand{\arraystretch}{1.2}
\begin{tabular}{|c||l||l|l|l||l|l|}
\cline{2-7} \multicolumn{1}{c||}{}
           & now & Run IIA & Run IIB & LHC & LC  & GigaZ \\ \hline  \hline
                                                                     
$\de\sweff (\times 10^5)$ & 17   & 50 & ~~13 & ~21 & (6)  & ~1.3  \\ \hline
$\de\MW$~[MeV]            & 37   & 30 & ~~15 & ~15 & 15   & ~6    \\ \hline
$\de\mt$~[GeV]            & ~5.1 & ~4 & ~~~2 & ~~2 & ~0.2 & ~0.13 \\ \hline
$\de\MH$~[MeV]            & ---  & ~--- & 2000 & 100 & 50   & 50    \\ \hline
\end{tabular}
\end{center}
\end{table}

In this talk, we discuss the potential impact of high precision measurements 
of $\sweff$, $\MW$, and other observables on the (indirect) determination 
of the Higgs boson mass in the SM and of non-SM mass scales in the MSSM. 
These unknown mass scales affect the predictions of the precision observables 
through loop corrections.


\section{Higgs Sector of the SM}

Within the SM, the precision observables measured at the $Z$~peak are affected 
by two high mass scales: the top quark mass, $\mt$, and the Higgs boson mass, 
$\MH$. They enter into various relations between electroweak observables. 
For example, the radiative corrections entering the relation between 
$\MW$ and $\MZ$, and between $\MZ$ and $\sweff$, have a strong quadratic 
dependence on $\mt$ and a logarithmic dependence on $\MH$. We mainly focus on 
the two electroweak observables that are expected to be measurable with 
the highest accuracy at GigaZ, $\MW$ and $\sweff$. The current theoretical
uncertainties~\cite{PCP} are dominated by the parametric uncertainties from 
the errors in the input parameters $\mt$ (see Table \ref{tab:precallcoll}) 
and $\Delta\alpha$. The latter denotes the QED-induced shift in the fine 
structure constant, $\al \to \al(\MZ)$, originating from charged-lepton and 
light-quark photon vacuum polarization diagrams. The hadronic contribution 
to $\De\al$ currently introduces an uncertainty of 
$\de\De\al = \pm 2 \times 10^{-4}$~\cite{delalphatheorydriven}. Forthcoming 
low-energy \epem annihilation experiments may reduce this uncertainty to about 
$\pm 5 \times 10^{-5}$~\cite{delalphajegerlehner}. Combining this value with 
future (indistinguishable) errors from unknown higher order corrections, we 
assign the total uncertainty of $\de\De\al = \pm 7 \times 10^{-5}$ to $\De\al$.
For the future theoretical uncertainties from unknown higher-order corrections
(including the uncertainties from $\de\De\al$) we assume,
\BE
\delta\MW(\mbox{theory}) = \pm 3 \mev, \quad
\delta\sweff(\mbox{theory}) = \pm 3 \times 10^{-5} \quad
\mbox{(future)} .
\label{eq:futureunc}
\EE
Given the high precision of GigaZ, also the experimental error in $\MZ$, 
$\de\MZ = \pm 2.1 \mev$, results in non-negligible uncertainties 
of $\de\MW = \pm 2.5 \mev$ and $\de\sweff = \pm 1.4 \times 10^{-5}$.
The experimental error in the top-quark mass, $\de\mt = \pm 130 \mev$, induces 
further uncertainties of $\de\MW = \pm 0.8 \mev$ and 
$\de\sweff = \pm 0.4 \times 10^{-5}$. Thus, while currently the experimental 
error in $\MZ$ can safely be neglected, for the GigaZ precision it will 
actually induce an uncertainty in the prediction of $\sweff$ that is larger 
than its experimental error.

\begin{itemize}
\item
The relation between $\sweff$ and $\MZ$ can be written as,
\BE
\label{deltars}
  \sweff \cweff = \frac{A^2}{\MZ^2 (1-\Delta r_Z)}, 
\EE
where $A = [(\pi\alpha)/(\sqrt{2} G_F)]^{1/2} = 37.2805(2)$~GeV is 
a combination of two precisely known low-energy coupling constants, the   
Fermi constant, $\gf$, and the electromagnetic fine structure constant, 
$\al$. The quantity $\Delta r_Z$ 
summarizes the loop corrections, which at the \onel\ level can be
decomposed as, 
\BE
\Delta r_Z  =  \Delta \al - \De\rho^{\rm t} +\De r_Z^{\rm H}
               + \cdots .
\label{deltarz}
\EE
The leading top contribution to the $\rho$ parameter~\cite{rhoparameter}, 
quadratic in $\mt$, reads,
\BE
\De \rho^{\rm t} = \frac{3 G_F m_t^2}{8\pi^2\sqrt{2}} \, .
\EE
The Higgs boson contribution is screened and logarithmic for large Higgs boson 
masses, 
\BE
\De r_Z^{\rm H}  =   \frac{G_F \MW^2}{8\pi^2\sqrt{2}} 
                       \frac{1+9\sw^2}{3\cw^2}
                   \log \frac{\MH^2}{\MW^2} +  \cdots .
\EE

\item
An independent analysis can be based on the precise measurement of $\MW$ near 
threshold. The $\MW$--$\MZ$ interdependence is given by,
\BEA
  \frac{\MW^2}{\MZ^2} 
  \left(1-\frac{\MW^2}{\MZ^2} \right) & = &
  \frac{A^2}{\MZ^2 (1-\Delta r)},
\label{eq:b}
\EEA
where the quantum correction $\Delta r$ has the \onel\ decomposition,
\BEA
\Delta r & = & \Delta \al - \frac{\cw^2}{\sw^2} \De\rho^{\rm t}
              + \De r^{\rm H} 
             + \cdots    , \\ 
\De r^{\rm H} & = & \frac{G_F \MW^2}{8\pi^2\sqrt{2}} \frac{11}{3}  
                  \log \frac{\MH^2}{\MW^2} +  \cdots ,
\EEA
with $\De\al$ and $\De\rho^{\rm t}$ as introduced above.

Due to the different dependences of $\sweff$ and $\MW$ on $\mt$ and $\MH$,
the high precision measurements of these quantities at GigaZ (and other 
observables entering a global analysis) can determine $\mt$ and $\MH$. 
The expected accuracy in the indirect determination of $\MH$ from the radiative
corrections within the SM is displayed in Fig.~\ref{fig:mhmt}. To obtain 
these contours, the error projections in Table~\ref{tab:precallcoll} are 
supplemented by central values equal to the current SM best fit values 
for the entire set of current high precision observables~\cite{Erler99A}. 
For the theoretical uncertainties, Eq.~(\ref{eq:futureunc}) is used, while the 
parametric uncertainties, such as from $\als$ and $\MZ$, are automatically
accounted for in the fits\footnote{All fit results in this Section were 
obtained using GAPP~\cite{GAPP}.}. The allowed bands in the $\mt$--$\MH$ plane 
for the GigaZ accuracy are shown separately for $\sweff$ and $\MW$. By adding 
the information on the top-quark mass, with $\delta\mt \lsim 130$~MeV 
obtained from measurements of the $t\bar{t}$ production cross section near
threshold, an accurate determination of the Higgs boson mass becomes feasible,
from both, $\MW$ and $\sweff$. If the two values are found to be consistent, 
they can be combined and compared to the Higgs boson mass measured in direct 
production through Higgs-strahlung~\cite{lep2higgs} (see the last row in 
Table~\ref{tab:precallcoll}). In Fig.~\ref{fig:mhmt} this is shown by 
the shaded area, where the measurements of other $Z$~boson properties as 
anticipated for GigaZ are also included. For comparison, the area corresponding
to current experimental accuracies is also shown.

\begin{figure}[htb!]
\vspace{-1em}
\begin{center}
\epsfig{figure=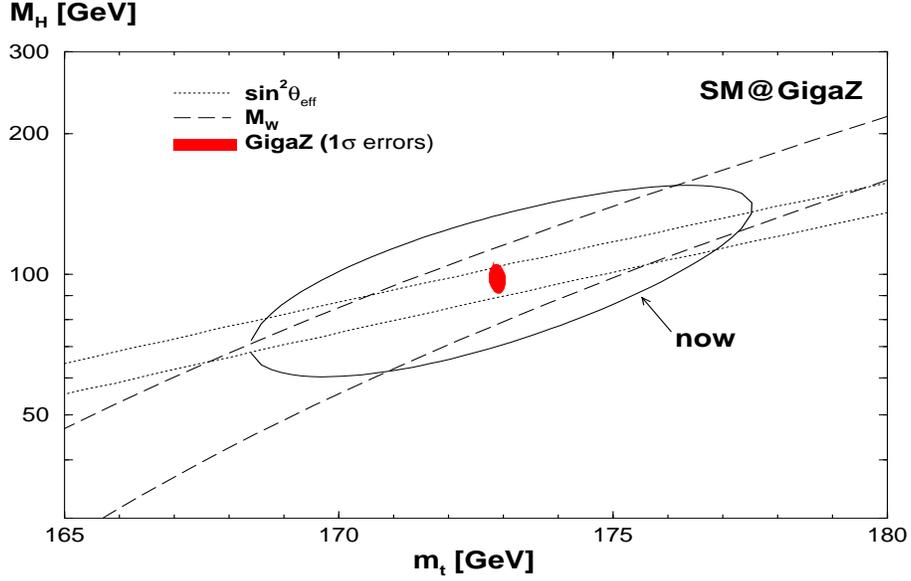,width=12cm,height=8cm}
\caption[0]{$1\sigma$ allowed regions in the $\mt$-$\MH$ plane taking 
into account the current measurements and the anticipated GigaZ 
precisions for $\sweff, \MW, \Ga_Z, R_l, R_q$ and $\mt$ (see text).}
\label{fig:mhmt}
\end{center}
\end{figure}
%
\begin{figure}[hb!]
\vspace{-4em}
\hspace{-2em}
\begin{center}
\epsfig{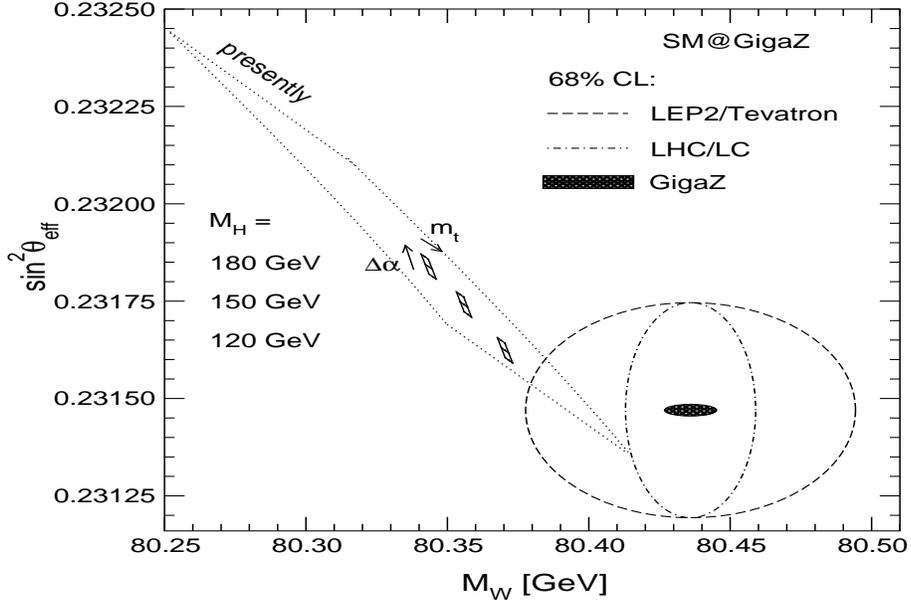}
\caption[0]{The theoretical prediction for the relation between $\sweff$ 
and $\MW$ in the SM for Higgs boson masses in the intermediate range is 
compared to the experimental accuracies at LEP~2/Tevatron (Run IIA), 
LHC/LC and GigaZ (see Table~\ref{tab:precallcoll}). For the theoretical 
prediction an uncertainty of $\de\De\al = \pm 7 \times 10^{-5}$ and 
$\de\mt = \pm 200 \mev$ is taken into account.}
\label{fig:SWMW}
\end{center}
\end{figure}

The results can be summarized by calculating the accuracy with which $\MH$ 
can be determined indirectly. The expectations for $\de\MH/\MH$ in each step
until GigaZ are collected in Table~\ref{tab:indirectmh}. It is apparent 
that GigaZ, reaching $\de\MH/\MH = \pm 7\%$, triples the precision in $\MH$ 
relative to the anticipated LHC status. On the other hand, a linear collider 
without the high luminosity option would provide only a modest improvement.

\def\tableline{\noalign{
\hrule height.7pt depth0pt\vskip3pt}}
\begin{table}[t!]
\caption{\label{tab:indirectmh}
{\sl Cumulative\/} expected precisions of indirect Higgs mass
determinations, given the error projections in Table \ref{tab:precallcoll}. 
Theoretical uncertainties and their correlated effects on $\MW$ and $\sweff$ 
are taken into account (see text).
The last column shows the indirect Higgs mass determination from the
full set of precision observables.}
\begin{center}
\setlength{\tabcolsep}{9pt}
\renewcommand{\arraystretch}{1.2}
\begin{tabular}{|l||r|r||r|}
\cline{2-4} \multicolumn{1}{c||}{}
                                       & $\MW$ & $\sweff$ & all  \\  \hline
                                                                     \hline
now                                    &200 \% & 62 \%    & 60 \% \\ \hline
                                                                     \hline
Tevatron Run IIA                       & 77 \% & 46 \%    & 41 \% \\ \hline
Tevatron Run IIB                       & 39 \% & 28 \%    & 26 \% \\ \hline
LHC                                    & 28 \% & 24 \%    & 21 \% \\ \hline
                                                                     \hline
LC                                     & 18 \% & 20 \%    & 15 \% \\ \hline
GigaZ                                  & 12 \% &  7 \%    &  7 \% \\ \hline
\end{tabular}
\end{center}
\end{table}

\item
A direct formal relation between $\MW$ and $\sweff$ can be established
by combining the two relations (\ref{deltars}) and (\ref{eq:b}) as,
\BE
\label{deltar}
\MW^2 = \frac{A^2}{\sweff (1 - \De r_W)} .
\EE
The quantum correction $\De r_W$ is independent of $\De\rho^{\rm t}$
in leading order and has the \onel\ decomposition,
\BEA
\De r_W  &=&  \De\al - \De r_W^{\rm H} + \cdots , \\
\De r_W^{\rm H}  &=&  \frac{G_F \MZ^2}{24\pi^2\sqrt{2}} 
                  \log \frac{\MH^2}{\MW^2} +  \cdots .
\label{deltarw}
\EEA
Relation~(\ref{deltar}) can be 
evaluated by inserting the measured value of the Higgs boson mass as 
predetermined at the LHC and the LC. This is visualized in Fig.~\ref{fig:SWMW},
where the present and future theoretical predictions for $\sweff$ and
$\MW$ (for different values of $\MH$) are compared with the experimental
accuracies at various colliders. Besides the independent predictions of
$\sweff$ and $\MW$ within the SM, the $\MW-\sweff$ contour plot in
Fig.~\ref{fig:SWMW} can be interpreted as an additional indirect
determination of $\MW$ from the measurement of $\sweff$.
Given the expected negligible error in $\MH$, this results in an
uncertainty of
\BE
\de\MW ({\rm indirect}) \approx \pm 2 \mev \pm 3 \mev .
\label{delmw2}
\EE
The first uncertainty reflects the experimental error in $\sweff$, while the 
second is the theoretical uncertainty discussed above (see
Eq.~(\ref{eq:futureunc})). The combined uncertainty of this indirect
prediction is about the same as the one of the SM
prediction according to Eq.~(\ref{eq:b})
and is close to the experimental error expected from the 
$W^+W^-$ threshold given in Eq.~(\ref{delmw}). 
\end{itemize}

Consistency of all the theoretical relations with the experimental data would 
be the ultimate precision test of the SM based on quantum fluctuations. 
The comparison between theory and experiment can also be exploited to constrain
possible physics scales beyond the SM. These additional contributions can 
conveniently be described in terms of the S,T,U~\cite{Peskin90} or $\epsilon$
parameters~\cite{Altarelli90}. Adopting the notation of Ref.~\cite{Erler99A}, 
the errors with which they can be measured at GigaZ are given as follows:
\BE
\label{eq:STU}
\begin{array}{lr}
  \Delta S = \pm 0.05, & \hspace{50pt} \Delta \hat{\epsilon}_3 = \pm 0.0004, \\
  \Delta T = \pm 0.06, & \hspace{50pt} \Delta \hat{\epsilon}_1 = \pm 0.0005, \\
  \Delta U = \pm 0.04, & \hspace{50pt} \Delta \hat{\epsilon}_2 = \pm 0.0004.
\end{array}
\EE
The oblique parameters in Eq.~(\ref{eq:STU}) are strongly correlated.
On the other hand, many types of new physics predict $U = \hat\epsilon_2 = 0$ 
or very small (see Ref.\cite{Erler99A} and references therein). 
With the $U$ ($\hat\epsilon_2$) parameter known, the anticipated 
errors in $S$ and $T$ would decrease to about $\pm 0.02$, while the errors
in $\hat\epsilon_1$ and $\hat\epsilon_3$ would be smaller than $\pm 0.0002$.


\section{Supersymmetry}

We now assume that supersymmetry would be discovered at LEP~2, the Tevatron, 
or the LHC, and further explored at an \epem\ linear collider. The high 
luminosity expected at TESLA can be exploited to determine supersymmetric 
particle masses and mixing angles with errors from \order{1\%} down to one per 
mille, provided they reside in the kinematical reach of the collider, which we 
assume to be about 1~TeV.

In contrast to the Higgs boson mass in the SM, the lightest $\cp$-even MSSM 
Higgs boson mass, $\Mh$, is not a free parameter but can be calculated from 
the other SUSY parameters. In the present analysis, the currently most precise 
result based on Feynman-diagrammatic methods~\cite{mhiggs2l} is used, relating 
$\Mh$ to the pseudoscalar Higgs boson mass, $\MA$. The numerical evaluation has
been performed with the Fortran code \fh~\cite{feynhiggs}. Later in
our analysis we also assume a future uncertainty in the theoretical
prediction of $\Mh$ of $\pm 0.5 \gev$.

The relation between $\MW$ and $\sweff$ is affected by the parameters of 
the supersymmetric sector, especially the $\Stop$-sector. At the LHC and in 
the first phase of LC operations, the mass of the light $\Stop$, $\mste$, and 
the $\Stop$-mixing angle, $\tst$, may be measurable very well, particularly in 
the process $e^+\,e^- \to \Stope \bar{\Stope}$ (see the last paper of
Ref.~\cite{gigaz} and references therein). On the other hand, background 
problems at the LHC and lacking energy at the LC may preclude the analysis of 
the heavy $\Stop$-particle,~$\Stopz$.

\begin{figure}[ht!]
\begin{center}
\epsfig{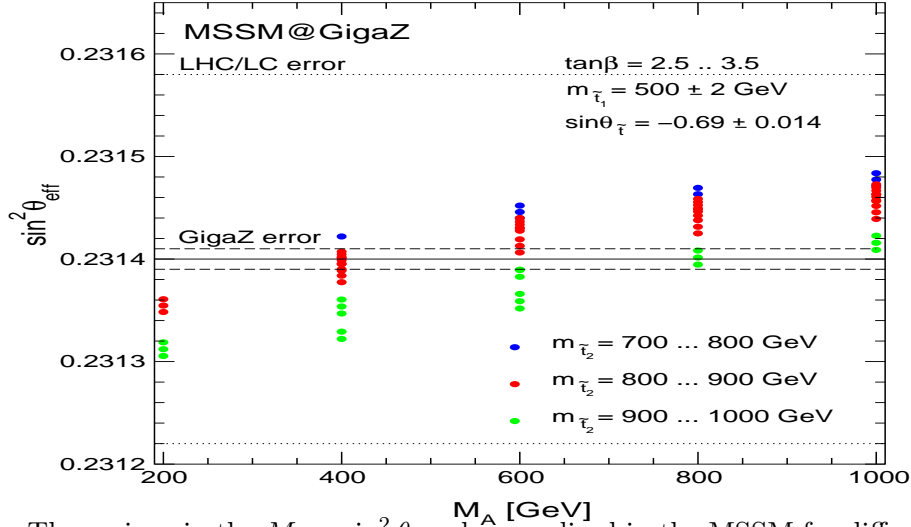}
\vspace{-1.5em}
\caption[0]{
The regions in the $\MA-\sweff$ plane realized in the MSSM for
different values of $\mstz$. The precision on $\sweff$ obtainable at the
LHC and the LC is compared to the prospective GigaZ precision around
the value $\sweff = 0.23140$, for $2.5 < \tb < 3.5$ and the other
experimental values as in 
Fig.~\ref{fig:MSt2MA}. (See text for details.)
}
\label{fig:sw2effMA}
\end{center}
\end{figure}
%
\begin{figure}[b!]
\vspace{-3em}
\begin{center}
\epsfig{figure=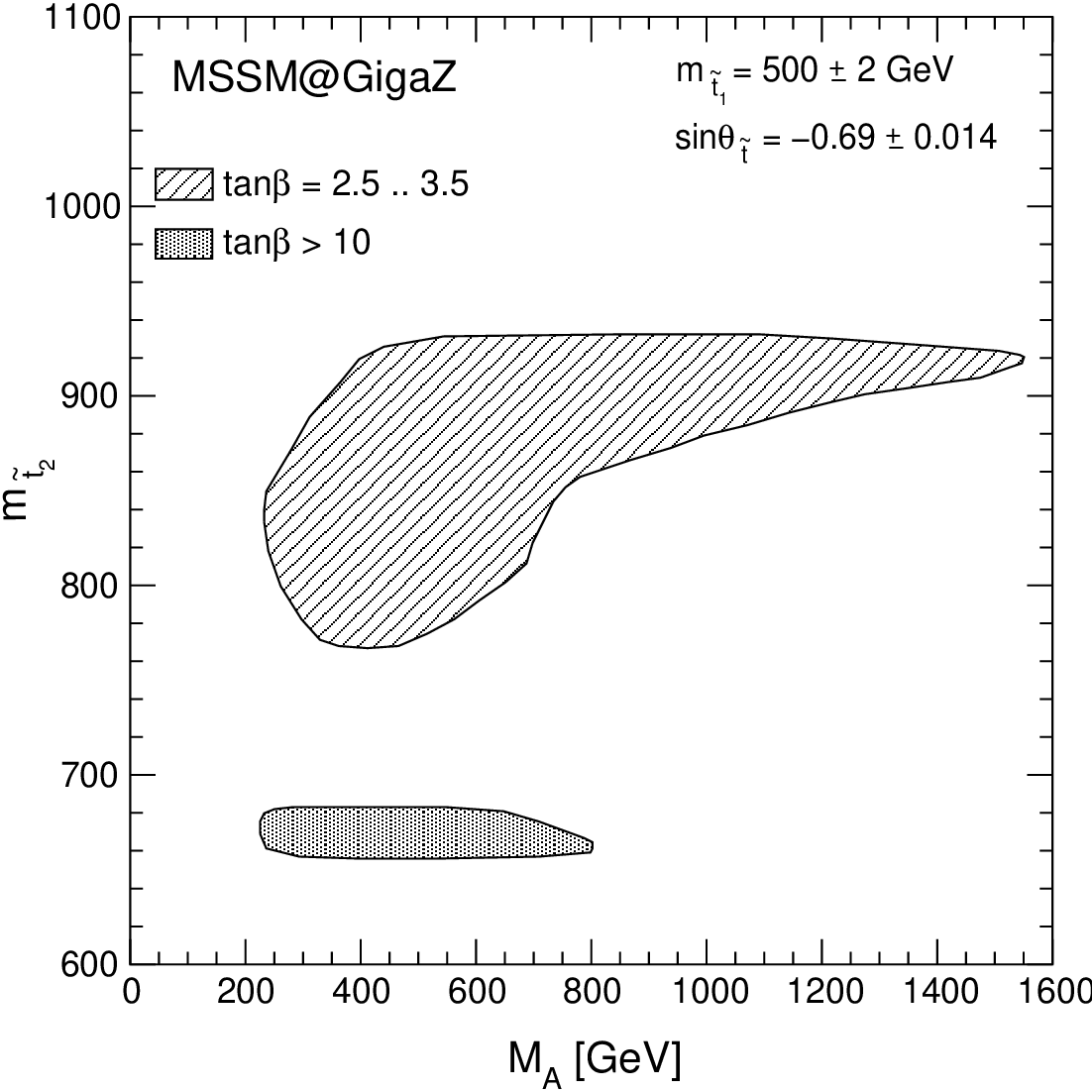,width=12cm,height=7.5cm}
\vspace{-1.5em}
\caption[0]{The region in the $\MA-\mstz$ plane, allowed by $1\,\si$ errors 
obtained from the GigaZ measurements of $\MW$ and $\sweff$: 
$\MW = 80.40 \gev$, 
$\sweff = 0.23140$, and from the LC measurement of $\Mh$: $\Mh = 115 \gev$. 
The experimental errors for the SM parameters are given in 
Table~\ref{tab:precallcoll}. 
$\tb$ is assumed to be experimentally constrained by $2.5 < \tb < 3.5$ or 
$\tb > 10$. The other parameters including their uncertainties are given by 
$\mste = 500 \pm 2 \gev$,
$\sintt = -0.69 \pm 2\%$,
$\Ab = \At \pm 10\%$,
$\mu = -200 \pm 1 \gev$,
$M_2 = 400 \pm 2 \gev$ and
$\mgl = 500 \pm 10 \gev$.
For the uncertainties of the theoretical predictions we use 
Eq.~(\ref{eq:futureunc}).}
\label{fig:MSt2MA}
\end{center}
\vspace{-1em}
\end{figure}

In Fig.~\ref{fig:sw2effMA} we show in a first step of the analysis the
effect of the precise determination of $\sweff$ alone on the indirect
determination of the heavier $\Stop$ mass, $\mstz$. In this first
step we neglect the variation of the SUSY parameters and the
theoretical uncertainty of $\Mh$. For the precision observables we
have taken $\sweff = 0.23140$ and $\Mh = 115 \gev$ with the experimental errors
given in the last column of Table~\ref{tab:precallcoll}. For $\tb$, the ratio 
of the vacuum expectation values of the two Higgs doublets in the MSSM,
we assume a relatively well determined $\tb = 3 \pm 0.5$, as can be expected 
from measurements in the gaugino sector (see e.g.\ Ref.~\cite{tbmeasurement}).
As for the other parameters, the following values are assumed:
$\mste = 500 \pm 2 \gev$,
$\sintt = -0.69 \pm 2\%$,
$\Ab = \At$,
$\mu = -200 \gev$,
$M_2 = 400 \gev$  and
$\mgl = 500 \gev$.
($A_{b,t}$ are trilinear soft SUSY-breaking parameters, $\mu$ is the Higgs 
mixing parameter, $M_2$ is one of the soft SUSY-breaking parameter in 
the gaugino sector, and $\mgl$ denotes the gluino mass.)
In Fig.~\ref{fig:sw2effMA} the GigaZ precision for $\sweff$ is
compared to the precision obtainable at the LHC and a LC without the
GigaZ option. While the LHC/LC precision gives no restrictions for
$\mstz$ or $\MA$, the high GigaZ precision could give lower and 
{\em upper} bounds on both non-SM mass parameters.

However, a more realistic scenario includes the other precision observable 
that can be determined at GigaZ with extremely high precision, $\MW$. 
In addition, all uncertainties of the additional SUSY mass scales, as well as 
the theoretical uncertainty of the Higgs boson mass prediction have 
to be taken into account. Therefore, as a second step in our analysis we now 
consider $\sweff$ and $\MW$ and include all possible uncertainties. It is
demonstrated in Fig.~\ref{fig:MSt2MA} how in this complete analysis limits on  
$\mstz$ and $\MA$ can be derived from measurements of $\Mh$, $\MW$, and 
$\sweff$. As experimental values we assumed $\Mh = 115 \gev$, 
$\MW = 80.40 \gev$, and $\sweff = 0.23140$, with the experimental errors 
given in the last column of Table~\ref{tab:precallcoll}, and the future 
theoretical uncertainty for the Higgs boson mass of $\pm 0.5 \gev$. 
We now consider two cases for $\tb$: the low 
$\tb$ region, where we assume a band, $2.5 < \tb < 3.5$ as for
Fig.~\ref{fig:sw2effMA}, and the high $\tb$ 
region where we assume a lower bound, $\tb \geq 10$ (see e.g.\
Ref.~\cite{tbmeasurement} and references therein). 
As for the other parameters, the following values are assumed, with 
uncertainties as expected from LHC and TESLA: 
$\mste = 500 \pm 2 \gev$,
$\sintt = -0.69 \pm 2\%$,
$\Ab = \At \pm 10\%$,
$\mu = -200 \pm 1 \gev$,
$M_2 = 400 \pm 2 \gev$  and
$\mgl = 500 \pm 10 \gev$.

In this full analysis, taking into account all possible uncertainties,
for low $\tb$ the heavier $\Stop$-mass, $\mstz$, can be restricted to 
$760 \gev \lsim \mstz \lsim 930 \gev$. The mass $\MA$ varies between 
$200 \gev$ and $1600 \gev$. A reduction of this interval to $\MA \ge 500 \gev$ 
by its non-observation at the LHC and the LC does not improve the bounds on 
$\mstz$. If $\tb \ge 10$, the allowed region turns out to be much smaller 
($660 \gev \lsim \mstz \lsim 680 \gev$), and $\MA$ is restricted to 
$\MA \lsim 800 \gev$. 

In deriving the bounds on $\mstz$, both the constraints 
from $\Mh$ (see Ref.~\cite{tampprec}) and $\sweff$ play an important role. 
For the bounds on $\MA$, the main effect comes from $\sweff$. We have assumed 
a value for $\sweff$ slightly different from the corresponding value obtained 
in the SM limit. For this value the (logarithmic) dependence on $\MA$
(see also Fig.~\ref{fig:sw2effMA}) is still 
large enough so that in combination with the high precision in $\sweff$ at 
GigaZ an {\em upper limit} on $\MA$ can be set. For an error as obtained at 
an LC without the GigaZ mode (see Table~\ref{tab:precallcoll}) no bound on 
$\MA$ could be inferred. Thus, the high precision measurements of $\MW$, 
$\sweff$, and $\Mh$ do not improve the direct lower bound on the mass of 
the pseudoscalar Higgs boson $A$, but instead they enable us to set an 
{\em upper bound}.


\section{Conclusions}

The opportunity to measure electroweak observables very precisely in the GigaZ 
mode of the projected \epem\ linear collider TESLA, in particular 
the electroweak mixing angle $\sweff$ and the $W$~boson mass, opens new areas 
for high precision tests of electroweak theories. We have analyzed in detail 
two examples: (i) The Higgs mass of the Standard Model can be extracted
to a precision of a few percent from loop corrections. By comparison with 
the direct measurements of the Higgs mass, bounds on new physics scales can be
inferred that may not be accessible directly. (ii) The masses of particles in 
supersymmetric theories, which for various reasons may not be accessible 
directly neither at the LHC nor at the LC, can be constrained. Typical 
examples are the heavy scalar top quark and the mass of the $\cp$-odd
Higgs boson, $\MA$.  (Further examples for the MSSM 
have also been studied in the original literature~\cite{gigaz}.)  
In the scenarios studied here, a sensitivity of up to order 2 TeV for the mass 
of the pseudoscalar Higgs boson and an upper bound of about 1 TeV for the heavy
scalar top quark can be expected. Opening windows to unexplored energy scales 
renders these analyses of virtual effects an important tool for experiments in 
the GigaZ mode of a future \epem\ linear collider.


\Acknowledgments
It is a pleasure for J.E.\ to thank the organizers for a very pleasant
conference in a spectacular setting. 


\end{document}